\documentclass[12pt,aps,preprint,nofootinbib]{revtex4}
\usepackage{graphicx,amsmath}
\usepackage{amssymb}
\usepackage{amsfonts}
\usepackage{hyperref}
\usepackage{color}
\usepackage{float}

\newcommand{\met}{\ensuremath{{\not\mathrel{E}}_T}}

\begin{document}
\title{Exotic Higgs Decay via $AZ/HZ$ Channel: a Snowmass Whitepaper}
 \author{Baradhwaj Coleppa, Felix Kling, Shufang Su}
\email {baradhwa@email.arizona.edu, kling@email.arizona.edu,shufang@physics.arizona.edu}

\affiliation {Department of Physics,
University of Arizona,
Tucson, AZ 85721, USA}

 \begin{abstract}
 
 While the conventional searches for the Higgs boson focus on its decays to Standard Model (SM) particles: $\gamma\gamma$, $ZZ$, $WW$, $bb$ and $\tau\tau$, in this study, we explore the decays  $A\rightarrow HZ$ or $H\rightarrow AZ$.   Such decays could appear in the exotic decay of the observed SM-like Higgs $h \rightarrow AZ$ with a light $A$, or for extra Higgses in beyond the SM new physics scenarios with an extended Higgs sector.   We study the exclusion bounds as well as discovery reach at the LHC for various combinations of $(m_A, m_H)$ for  the process:  $gg\rightarrow A/H \rightarrow HZ/AZ \rightarrow bb\ell \ell$ for $\ell =e$, $\mu$.  We found that for 14 TeV LHC with 300 ${\rm fb}^{-1}$ integrated luminosity, the 95\% C.L. limits on $\sigma \times BR(gg\rightarrow A/H \rightarrow HZ/AZ \rightarrow bb\ell \ell)$ vary between about 30 fb to a few fb   for the parent heavy Higgs mass in the range  200 GeV to 600 GeV, while the limits for 5$\sigma$ discovery are about 2$-$3 times larger. Comparing 
with the specific case of Next to Minimal Supersymmetric Standard Model, we find that  this channel could be useful for heavy Higgs searches, especially for Higgs masses below 350 GeV. 
 
 \end{abstract}
\maketitle

\section{Introduction}
\label{sec:intro}

The discovery of the Standard Model (SM)-like Higgs at the CERN Large Hadron Collider (LHC)  provides the last link towards validating the SM \cite{Aad:2012oxa, Aad:2012gxa, CMS:2011opa, CMS:2013haz}.  However,   much work still remains in establishing the nature of the Higgs boson, as well as studying possible extensions of the SM Higgs sector. A complete characterization of all possible decay modes of the SM-like Higgs is crucial. While the discovery of the Higgs relied on conventional search channels $\gamma\gamma$, $ZZ$,    $WW$, $bb$ and $\tau\tau$ with event rates compatible with SM predictions, the limit on the branching ratio to exotic states is rather weak \cite{ATLAS:2013sla, CMS:yva}.  If the Higgs decay branching fractions indeed show a departure from the SM, it would become very important to get a handle on possible exotic decays as they will help narrow down the various possibilities beyond the SM.   A particularly interesting example is the exotic decay of the observed SM-like Higgs $h \
rightarrow AZ$ with a light CP-odd Higgs $A$.

Many beyond the SM scenarios are constructed by extending the Higgs sector - well known examples are the Minimal Supersymmetric Standard Model (MSSM) \cite{Nilles:1983ge,Haber:1984rc,Barbieri:1987xf}, Next to Minimal Supersymmetric Standard Model (NMSSM) \cite{Ellis:1988er,Drees:1988fc} and Two-Higgs Doublet Models (2HDM)\cite{Branco:2011iw,type1,hallwise,type2}.  In addition to the SM-like Higgs boson in these models, there are other CP-even Higgses, CP-odd Higgses, as well as charged ones.  Other than the decay of those extra Higgses into the SM-final states $\gamma\gamma$, $ZZ$,    $WW$, $bb$ and $\tau\tau$, which have been the focus of the current Higgs searches, the decay of heavy Higgses into light Higgses, or Higgs plus gauge boson could also be sizable.  In this study, we explore the decay of  $H\rightarrow AZ$ or $A\rightarrow HZ$, with $H$ and $A$ referring to the CP-even and CP-odd Higgs respectively.  In generic 2HDM or NMSSM, both decays  $H_i\rightarrow A_jZ$ 
and $A_i\rightarrow H_jZ$ could appear with large branching fractions~\cite{Craig:2012vn, Chiang:2013ixa, Christensen:2013dra, Grinstein:2013npa}. $A\rightarrow h^0 Z$ could also have a sizable cross-section in the  low $\tan\beta$ region   of the MSSM with the light CP-even $h^0$ being SM-like \cite{Lewis:2013fua}.

In this work, we study the process  $gg\rightarrow A/H \rightarrow HZ/AZ \rightarrow bb\ell \ell$ for $\ell =e$, $\mu$, exploring the exclusion bounds as well as discovery reach at the LHC for various combinations of $(m_A, m_H)$.   
  For simplicity, we use $A\rightarrow HZ$ to refer to both types of decays.  Since we do not make use of angular correlations, the results obtained for $A\rightarrow HZ$ apply to  $H\rightarrow AZ$  as well with the values of $m_A$ and $m_H$ switched.  Details of our analysis can be found in Ref.~\cite{AZ_detail}.  A similar analysis of $A\rightarrow H_{\rm SM}Z$ in the 2HDM has also been carried out in Ref.~\cite{AZ_2HDM}.

The paper is organized as follows.  In Sec.~\ref{sec:analysis}, we briefly present our analysis describing the efficiencies of the cuts employed for both the signal and dominant SM backgrounds.  In Sec.~\ref{sec:results}, we present the 95\% C.L. exclusion as well as 5$\sigma$ discovery limits for $\sigma \times BR(gg \rightarrow A \rightarrow HZ \rightarrow bb \ell \ell)$ at the 14 TeV LHC with 100, 300 and 1000 ${\rm fb}^{-1}$ integrated luminosity.   To make connection with existing models, we also compare these exclusion and discovery limits to the case of the NMSSM. We conclude in Sec.~\ref{sec:conclusions}.


 
 \section{Analysis}
 \label{sec:analysis}



We study the dominant gluon fusion Higgs production $gg \rightarrow A \rightarrow HZ$ with  $H \rightarrow b {b}$ and $Z \rightarrow \ell \ell$ for $\ell = e, \mu$. 
The dominant SM backgrounds for $bb \ell \ell $ final states are $Z/\gamma^*bb$ with leptonic $Z$ decay, $t\bar{t}$ with leptonically decaying top quarks, $ZZ \rightarrow bb\ell \ell $, and $H_{\rm SM}Z$~\cite{Cordero:2009kv, Kidonakis:2012db,Campbell:2011bn,Dittmaier:2011ti}.  We have ignored the subdominant backgrounds from $WZ$, $WW$, $H_{\rm SM} \rightarrow ZZ$, $Wbb$, Multijet QCD Background, $Zjj$, $Z\ell\ell$ as well as  $tWb$.  Those backgrounds either have small production cross sections, or can be sufficiently suppressed by the cuts imposed. 



We use Madgraph 5/MadEvent v1.5.11 \cite{Alwall:2011uj} to generate our signal and background events.  These
events are passed to Pythia v2.1.21 \cite{PYTHIA}  to simulate initial and final state radiation,  showering and
hadronization. The events are further passed through Delphes 3.07 \cite{Ovyn:2009tx} with the Snowmass combined LHC detector card \cite{snowmassdetector} to simulate detector effects.  

For the signal process, we generated event samples at 14 TeV LHC for $gg \rightarrow A \rightarrow HZ$  with the daughter particle mass fixed at $m_H=$200, 125, and 50   GeV while varying the parent particle mass $m_A$ in the range 150 $-$ 600 GeV.  We applied the following cuts to identify the signal from the backgrounds\footnote{Requiring the missing transverse energy  to be small would potentially greatly reduce the $t\bar{t}$ background.  However, including  pile-up effects introduces $\met $ in the signal events, which renders the cut inefficient.     We thank Meenakshi Narain and John Stupak for pointing this out to us. }: 
\begin{itemize}
	\item \textbf{Two leptons, Two tagged $b$'s:} We require exactly two identified isolated leptons and two tagged bottom jets using the Snowmass Delphes detector card: 
\begin{equation}
\begin{aligned}
n_{\ell}=2,  \quad \quad n_{b} =2 .
\end{aligned}
\label{cut_1}
\end{equation}
This implies that the leptons and bottom jets are in the central detector region $|\eta_{\ell,b} |<2.5$ and satisfy a minimum transverse momentum cut of $p_{T,\ell} > 10$ GeV and $p_{T,b}>$ 15 GeV. For jet reconstruction, the anti-$k_T$ jet algorithm with $R$=0.5 is used. 

	\item \textbf{Lepton Trigger:} We require minimum transverse momenta for the leptons to pass the trigger:
\begin{equation}
\begin{aligned}
 p_{T,\ell_1}>30 \text{ GeV}\ \text{or} \ p_{T,\ell_1}>20 \text{ GeV},\ p_{T,\ell_2}>10 \text{ GeV}.
\end{aligned}
\label{cut_2}
\end{equation}

	\item \textbf{Dilepton mass $m_{\ell \ell}$:} We require the dilepton mass to be in the $Z$-mass window:
\begin{equation}
\begin{aligned}
80 \text{ GeV} <m_{\ell \ell}<  100 \text{ GeV}.
\end{aligned}
\label{cut_4}
\end{equation}		
	
	\item \textbf{$m_{bb}$ vs. $m_{bb\ell \ell}$:} We require the dijet mass $m_{bb}$ to be close to the daughter-Higgs mass $m_H$ and the mass $m_{bb\ell\ell}$ to be close to the parent-Higgs mass $m_{A}$. The two masses are correlated, i.e., if we underestimate $m_{bb}$ we also underestimate $m_{bb\ell \ell}$. To take this into account we apply a two-dimensional cut: 
\begin{equation}
\begin{aligned}
(0.95 - w_{bb})\cdot m_{H} < \; &m_{bb} < (0.95 + w_{bb})\cdot m_{H} \text{ with } w_{bb}= 0.15; \\
\frac{m_Z + m_H}{m_A}\cdot(m_{bb\ell\ell}-  m_A - w_{bb\ell \ell} )< m_{bb} & -m_H < \frac{m_Z + m_H}{m_A}\cdot(m_{bb\ell\ell } -m_A + w_{bb\ell \ell} ), 
\end{aligned}
\label{cut_5}
\end{equation}	
	where $w_{bb}$ is the width for the dijet mass window. Note that the slightly shifted reconstructed Higgs mass $m_{bb}$ (0.95 $m_H$ instead of $m_H$) is due to the reconstruction of the $b$-jet with a small size of $R$=0.5. The second condition describes two lines going through the points $(m_A \pm w_{bb\ell \ell }, m_H)$ with slope $(m_Z + m_H)/m_A$. We choose a width for the $m_{bb\ell \ell }$ peak of $w_{bb\ell \ell } = \text{Max}(w_{SM}|_{m_A},0.075 m_A)$ where $w_{SM}|_{m_A}$ is the width of a SM Higgs with mass $m_A$. This accounts for both small Higgs masses for which the width of the peak is caused by detector effects and large Higgs masses for which the physical width dominates.   The effectiveness of this cut is shown in Fig.~\ref{fig:2D_mbb_mllbb} for $m_A=300$ GeV and $m_{H}=125$ GeV, with two horizontal lines indicating the $m_{bb}$ range and two slanted lines indicating the $m_{bb\ell\ell}$ range as given in Eq.~(\ref{cut_5}).
 
 \begin{figure}[h!]
	\includegraphics[scale=0.27]{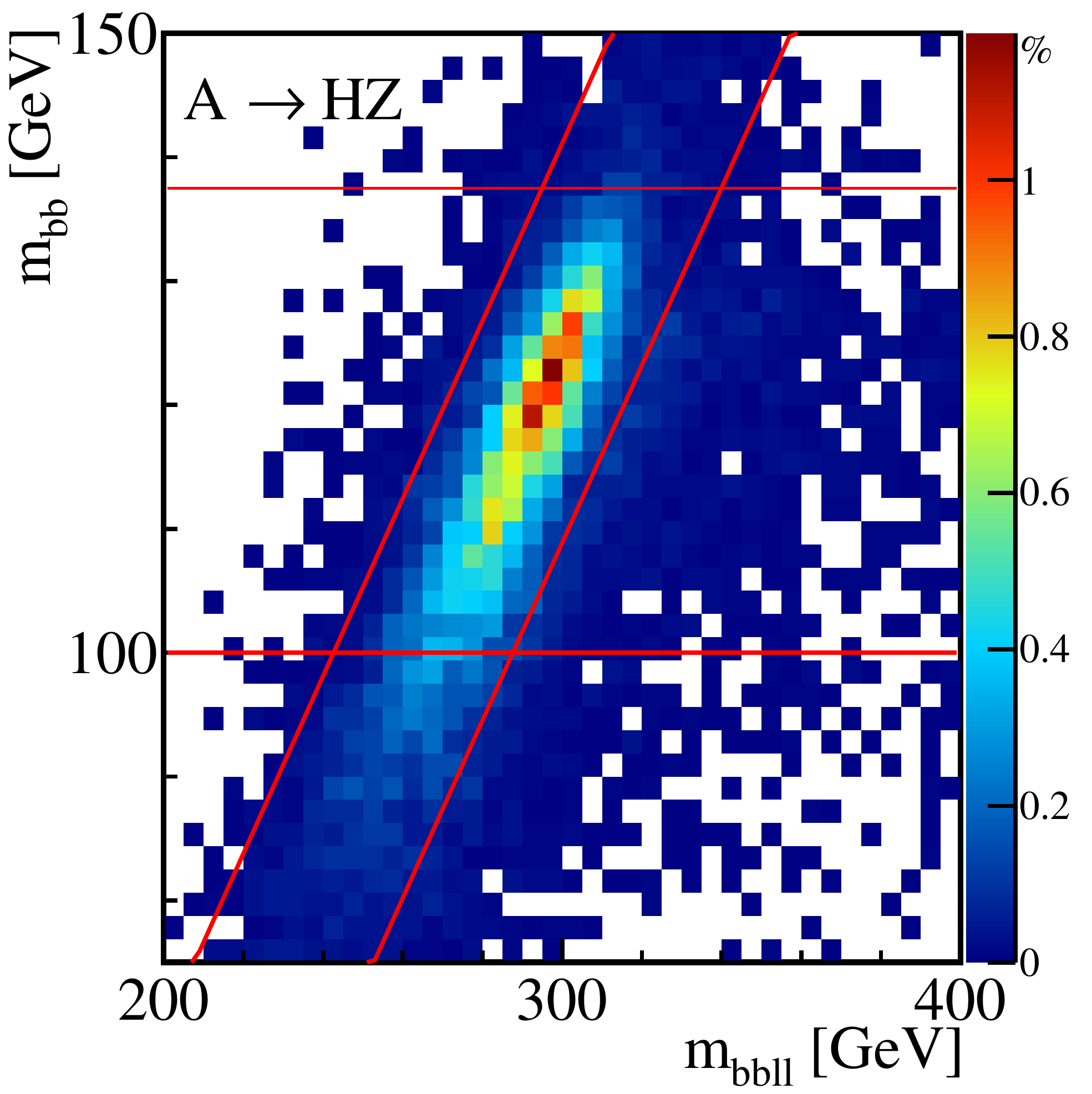}
	\includegraphics[scale=0.27]{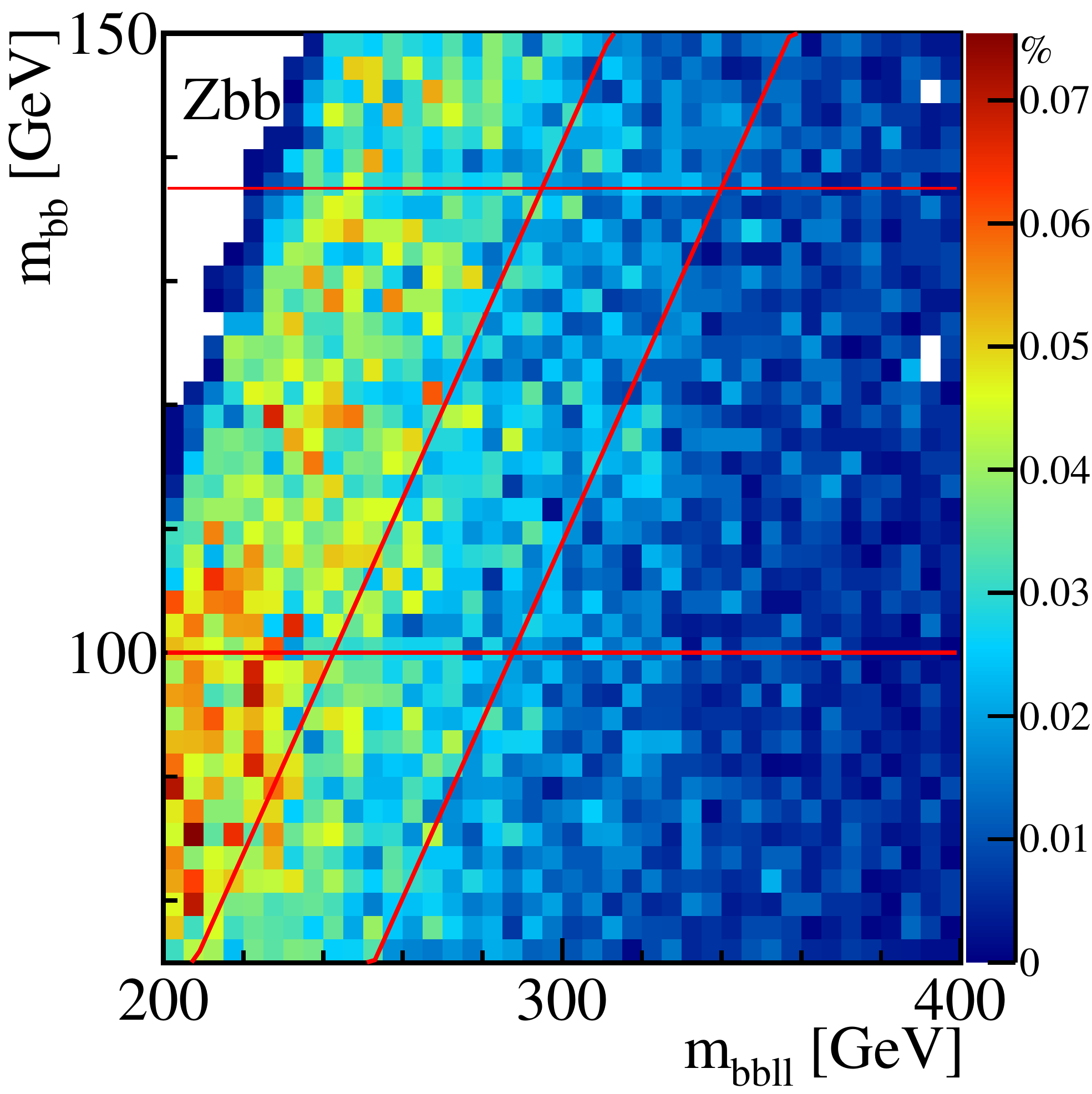}
	\includegraphics[scale=0.27]{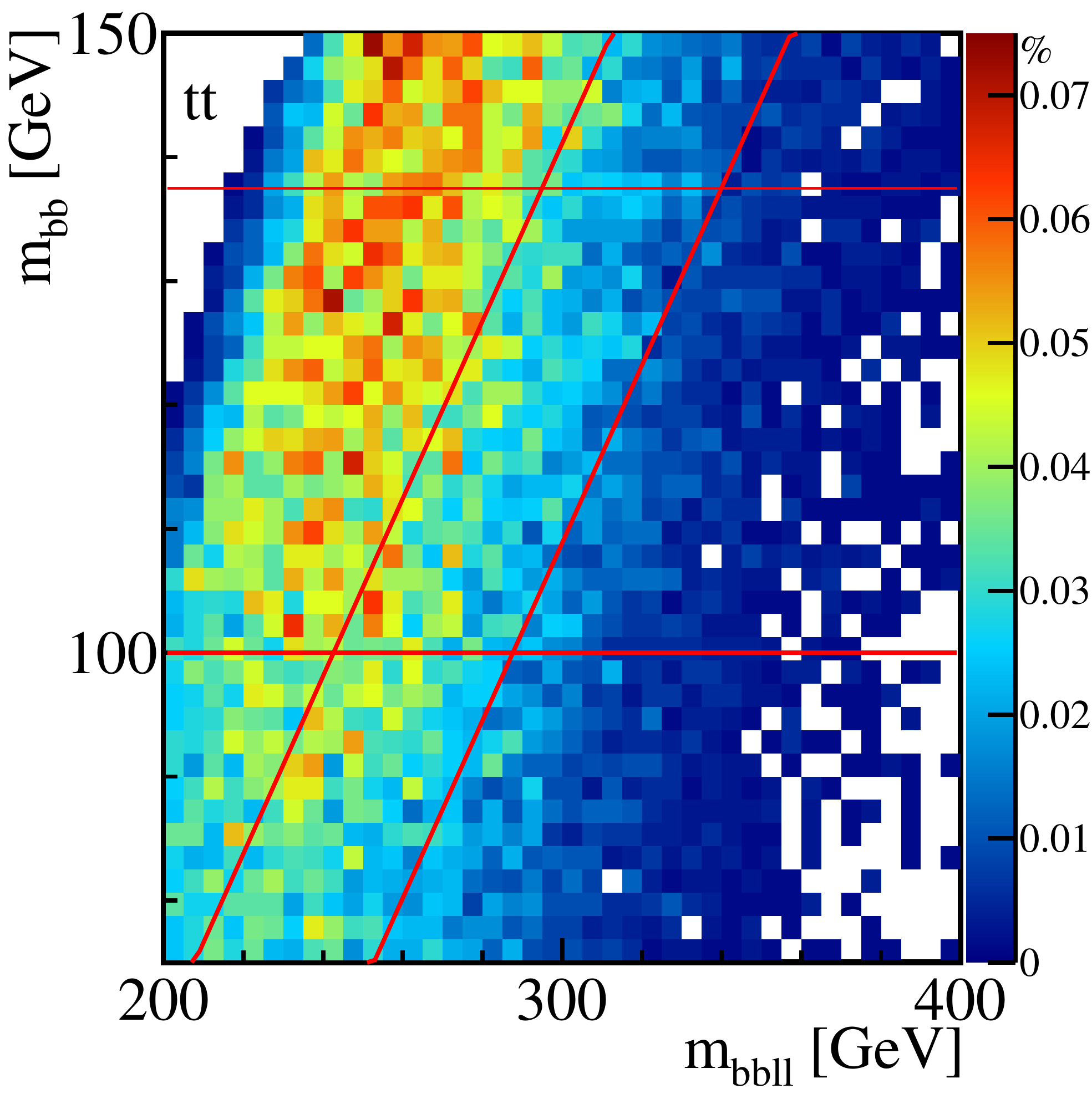}
\caption{Normalized distribution (in percent as given by the color code along the $y$-axis)  of $m_{bb}$ vs. $m_{bb\ell \ell }$ for the signal (left), and the dominating $Zbb$ (central) and $t\bar{t}$ (right) backgrounds   for $m_A=300$ GeV and $m_{H}=125$ GeV.  Two horizontal lines indicate the $m_{bb}$ range and two slanted lines indicate the $m_{bb\ell\ell}$ range,  as given in Eq.~(\ref{cut_5}).  }
\label{fig:2D_mbb_mllbb}
\end{figure}

	\item \textbf{Transverse momentum:} We require the sum of the transverse momenta of the bottom jets and the sum of the transverse momenta of the bottom jets and leptons to be large:
	\begin{equation}
\begin{aligned}
\sum_{b\, \rm{jets}}p_T&>\, 0.6 \cdot \frac{m_A^2+ m_H^2 - m_Z^2}{2 m_A} \\
\sum_{\ell,\, b\,\rm{jets}}p_T&>\, 0.66 \cdot  m_A 
\end{aligned}
\label{cut_6}
\end{equation}	

\end{itemize}
	



\begin{table}[htbp]
\centering
\begin{tabular}{l|r|rrrrrr}
\hline
Cut 							&Signal [fb]	&$bb\ell \ell$ [fb]	&$H_{\rm SM}Z$ [fb]	& $t\overline{t}$ [fb] 	&$S/B$		&$S/\sqrt{B}$	\\
\hline
$\sigma_{total}$					&  		&2.21$\cdot10^6$	&883			&9.20$\cdot10^5$	&-		&-		\\
leptonic decay						&100		&2.21$\cdot10^6$	&59.4			&2.15$\cdot10^4$	&-		&-		\\
Two leptons, Two $b$'s 		[Eq.(\ref{cut_1})]	&6.35 		&343		  	&3.44 			&1409	          	&0.0036 	&2.63		\\
Lepton trigger  		[Eq.(\ref{cut_2})]	&6.35 		&336		  	&3.44 			&1394           	&0.0037 	&2.65		\\
$m_{\ell \ell }$		[Eq.(\ref{cut_4})]	&5.76		&285			&3.13			&189			&0.012 		&4.59		\\
$m_{bb}$ vs $m_{bb\ell \ell}$	[Eq.(\ref{cut_5})]	&3.03 		&11.5			&0.401			&11.5			&0.14 		&11.5		\\
$\sum p_{T,b}$, $(\sum p_{T,b}+p_{T,\ell})$	[Eq.(\ref{cut_6})]	&2.81 	&8.11		&0.361			&8.38			&0.17 		&12.0		\\
\hline
\end{tabular}
\caption{Signal and background cross sections with cuts for the signal benchmark point $m_{A}$ = 300 GeV and $m_H$ = 125 GeV at the 14 TeV LHC.  We have chosen a nominal value for $\sigma \times BR(gg \rightarrow A \rightarrow HZ \rightarrow bb \ell \ell)$ of 100 fb to illustrate the cut efficiencies for the signal process.  The last column of $S/\sqrt{B}$ is shown for an integrated luminosity of ${\cal L}=300\  {\rm fb}^{-1}$.  }
\label{tab:cuts}
\end{table}

In Table \ref{tab:cuts}, we show the signal and background cross sections with cuts for signal benchmark point of $m_{A}$ = 300 GeV and $m_H$ = 125 GeV at the 14 TeV LHC.  We have chosen a nominal value for $\sigma \times BR(gg \rightarrow A \rightarrow HZ \rightarrow bb \ell \ell)$ of 100 fb to illustrate the cut efficiencies for the signal process.   The last column of $S/\sqrt{B}$ is shown for an integrated luminosity of ${\cal L}=300\  {\rm fb}^{-1}$. 

\section{Discovery and Exclusion Limits} 
\label{sec:results}

\begin{figure}[h!]
\includegraphics[scale=0.40]{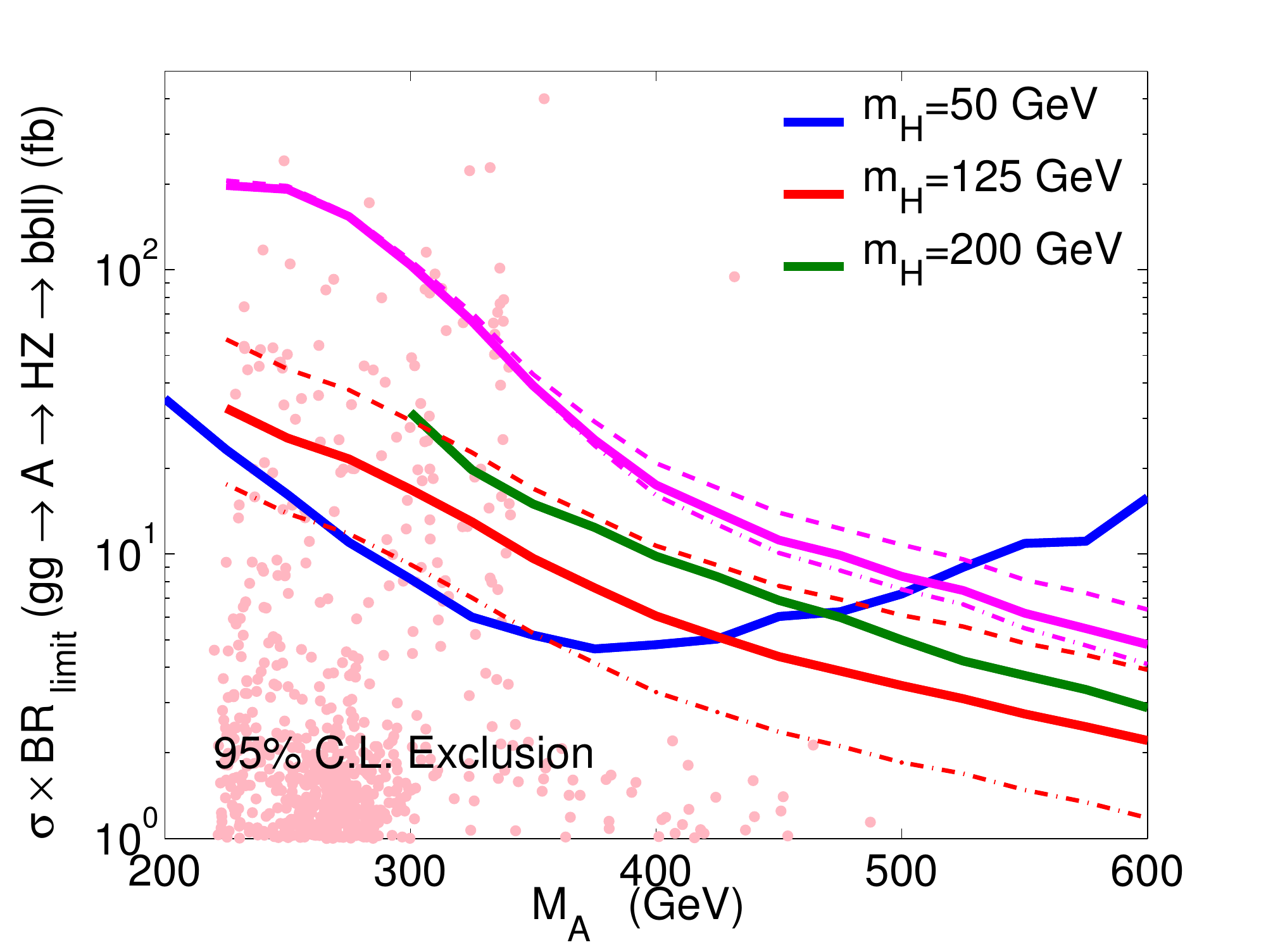}
\includegraphics[scale=0.40]{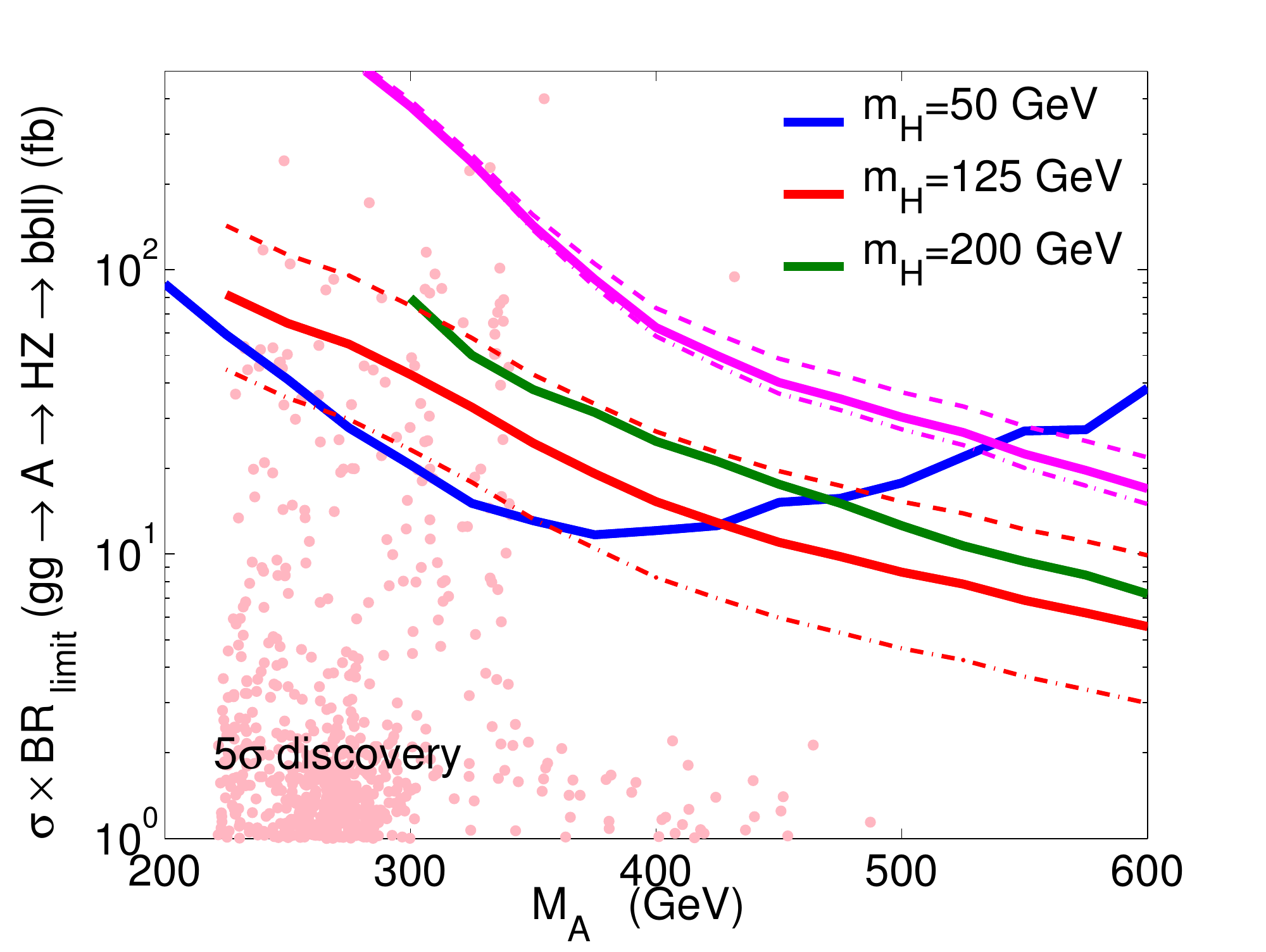}
 \caption{ 95\% C.L. exclusion (left panel) and 5$\sigma$ reach (right panel)  in $\sigma \times BR(gg \rightarrow A \rightarrow HZ \rightarrow bb \ell \ell)$ at the 14 TeV LHC with 300 ${\rm fb}^{-1}$ integrated luminosity as a function of the parent particle mass $m_A$ for various daughter particle mass $m_H=$ 50, 125 and 200 GeV.  Also shown are the reach of 100 ${\rm fb}^{-1}$ (dashed lines) and 1000 ${\rm fb}^{-1}$ (dash-dotted lines) for the case of $m_H=125$ GeV, as well as the reach  assuming 10\% systematic error for the backgrounds (purple lines).  Scattering dots are the possible $\sigma \times BR$ range  in the NMSSM for the daughter Higgs mass being $125\pm2$ GeV.}
\label{fig:cslimit}
\end{figure}

 In Fig.~\ref{fig:cslimit}, we show the 95\% C.L. exclusion (left panel) and 5$\sigma$ discovery (right panel) reach of $\sigma \times BR(gg \rightarrow A \rightarrow HZ \rightarrow bb \ell \ell)$ at the 14 TeV LHC with 300 ${\rm fb}^{-1}$ integrated luminosity  as a function of $m_A$ for $m_H=$ 50, 125 and 200 GeV.   Also shown are the reach of 100 ${\rm fb}^{-1}$ (dashed lines) and 1000 ${\rm fb}^{-1}$ (dash-dotted lines) for the case of $m_H=125$ GeV, as well as the reach assuming 10\% systematic error  for the backgrounds (purple lines).    The exclusion/discovery limits were calculated using the theta-auto program \cite{thetaauto}.   For 300 ${\rm fb}^{-1}$ integrated luminosity with $m_H=125$ GeV, the 95\% C.L. limit is about 26 (2.2)  fb for $m_A=250 \,(600)$ GeV, while the 5 $\sigma$ discovery limit is about 65 (5.6) fb.  The limits typically get better for large $m_A$ and smaller $m_H$, except for very large $m_A$ with very small $m_H$, when the two $b$ jets are highly collimated and thus do not 
pass the two $b$ jets requirement.


 Typical range of $\sigma \times BR$  in the NMSSM \cite{Christensen:2013dra} are plotted as scattered dots  for the daughter Higgs mass being $125\pm2$ GeV, which quickly drops for Higgs mass around 350 GeV due to the opening of the $t\bar{t}$ channel.   
 $A\rightarrow HZ$, or $H\rightarrow AZ$, with $bb \ell \ell$ final states, therefore, could be a useful channel for heavy Higgs searches, especially for Higgs mass below 350 GeV.   In the generic Type-II 2HDM,   $A/H\rightarrow H/AZ$ could even compete with the $t\bar{t}$ channel for Higgs mass above 350 GeV \cite{Coleppa:2013dya}.

\section{Conclusion}
\label{sec:conclusions}

While the conventional searches for the Higgs boson focus on its decays to SM final states: $\gamma\gamma$, $ZZ$, $WW$, $bb$ and $\tau\tau$, in this study, we explored the decay  $H\rightarrow AZ$ or $A\rightarrow HZ$ with $bb \ell\ell  $ final state and presented model independent limits on the 95\% C.L. exclusion and $5\sigma$ discovery   of the heavy Higgs in this channel.  For 14 TeV LHC with 300 ${\rm fb}^{-1}$ integrated luminosity, the 95\% C.L. limits on $\sigma \times BR$ vary between   30 fb  to a few fb for the parent heavy Higgs mass in the range of  200 GeV to 600 GeV, while the limit for 5$\sigma$ discovery is about 2$-$3 times larger.   Comparing with the specific case of NMSSM,   the $HZ/AZ$ channel can be   useful for Higgs masses all the way up to 350 GeV, when the $t\bar{t}$ channel opens up.   The reach could be even better in the generic 2HDM.



\begin{acknowledgements}
We thank Nathaniel Craig, Tao Han, Meenakshi Narain, Peter Loch, and John Stupak for helpful discussions.  This work was supported by  the Department of Energy under  Grant~DE-FG02-13ER41976.
\end{acknowledgements}

\end{document}